# Large Rashba parameter for *4d* strongly correlated perovskite oxide SrNbO$_3$ ultrathin films

Hikaru Okuma*, Yumiko Katayama, and Kazunori Ueno

Department of Basic Science, University of Tokyo,

Meguro, Tokyo 153-8902, Japan

*hikaruokuma613@g.ecc.u-tokyo.ac.jp

## Abstract

To elucidate the spin relaxation mechanism of SrNbO$_3$ (SNO) ultrathin films, the transport properties of a series of SNO films with various thicknesses $t$ were measured on both sides of the metal-insulator transition. The spin-orbit scattering time ($\tau_{so}$) was deduced from the analysis of the magnetoresistance with weak antilocalization theory, and it was found that $\tau_{so}$ was inversely proportional to the momentum scattering time. This result was explained in terms of the D'Yakonov-Perel' mechanism, indicative of the dominant Rashba effect. The values of the Rashba parameter were on the order of $1\times10^{-12}$ eVm, which were largest in the values reported for other ultrathin films of metallic oxides.

Electron systems lacking inversion symmetry show Rashba-type spin-orbit coupling (SOC) [1] [2]. Heterointerfaces with large Rashba SOCs [3] [4] [5] [6] have attracted considerable attention due to their potential spintronics applications and intriguing phenomena, such as the intrinsic spin Hall effect [7]. For spintronics, semiconductor heterointerfaces with large Rashba SOCs have been used to control spin precession in spin field effect transistors [8] [9]. Recently, a remarkably high spin-to-charge conversion rate [10] was reported for an oxide heterointerface with a large Rashba SOC, which is much larger than those of other Rashba systems, such as an interface between Ag and Bi (Ag/Bi) [11], a surface of a topological insulator $\alpha$-Sn [12], and heavy metals [13]. Among all systems with Rashba SOCs, oxide materials were air stable and easy to fabricate, but their Rashba parameters $\alpha_R$ were smaller than those of other systems with heavy elements. There has been a lot of effort spent finding oxide semiconductor heterostructures with large $\alpha_R$, such as $SrTiO_3$ (STO) [14] [15] [16] and $KTaO_3$ (KTO) [17] [18] based interfaces, all of which use high mobility oxide semiconductors STO or KTO. Furthermore, there have been several reports on $\alpha_R$ for metallic oxide ultrathin films, such as $SrIrO_3$ (SIO) [19] and $La_{2/3}Sr_{1/3}MnO_3$ (LSMO) [20], although $\alpha_R$ was even smaller for these materials than for semiconductor heterostructures. Ultrathin films are advantageous over interfaces because information about the electronic structure and local density of states can be directly obtained by using angle-resolved photoemission and scanning tunneling spectroscopy, respectively.

$SrNbO_3$ (SNO) has the same perovskite-type structure as STO but with a heavier transition metal Nb and a $4d^1$ electronic configuration for niobium, thus exhibiting metallic behavior. Bulk SNO is not stable under ambient conditions. Therefore, recently, SNO thin films were deposited on perovskite substrates in vacuo and have been intensively studied [21] [22] [23]. Since $Sr(Ti_{1-x}Nb_x)O_3$ (x=0.02–0.2) showed a systematic enhancement in the SOC strength with increasing Nb concentration [24] [25], SNO is a potential candidate for a Rashba interface with a high $\alpha_R$. However, two important issues must be addressed; one issue is that SNO films on semiconductor STO or KTO substrates could show parallel conduction between the SNO film and an oxide-deficient surface of the substrate. Since the doped substrate surface shows high mobility at low temperature, it is difficult to separate the intrinsic SNO and the substrate contributions. Another issue is that the study on $Sr(Ti_{1-x}Nb_x)O_3$ assumed a D'Yakonov-Perel' (DP)-type spin relaxation mechanism to estimate $\alpha_R$ [25], but there are other possible spin relaxation mechanisms. Thus, we must confirm that the DP-type mechanism dominates the spin relaxation in the SNO film.

Three types of spin relaxation mechanisms have been found for conductive electrons: Elliot-Yafet (EY) type, DP type and Bir-Aronov-Pikus (BAP) type [26]. The EY mechanism describes the spin inversion caused by the momentum scattering of an electron from impurities and phonons. This spin inversion originates from an entanglement of spin-up and spin-down states caused by the SOC of lattice ions. Therefore, the spin relaxation time ($\tau_{so}$) is proportional to the momentum scattering time ($\tau_p$). The DP mechanism originates from spin precession processes with elastic scattering events. Rashba SOC lifts the spin degeneracy, and then spin-up and spin-down bands have different energies. This is equivalent to having a momentum-dependent internal magnetic field, by which spin processes occur. Momentum scattering causes fluctuations in the internal magnetic field and interrupts spin precession. Therefore, $\tau_{so}$ is inversely proportional to $\tau_p$, in contrast to the behavior of the EY mechanism. The BAP mechanism originates from an electron–hole exchange interaction, while SNO is a $d^1$ metal and only has electrons on the Fermi surface. Thus, this mechanism can be disregarded for SNO films.

In this study, we examine the spin relaxation mechanism of SNO ultrathin films deposited on an insulator substrate. From the analysis of the magnetoresistance (MR) with weak antilocalization (WAL) theory for a series of films with different thicknesses (*t*) and disorder, we find that $\tau_{so}$ is inversely proportional to $\tau_p$, indicating DP-type spin relaxation and a dominant Rashba effect. The values of $\alpha_R$ are on the order of $1\times10^{-12}$ eVm, which are larger than the values reported for other ultrathin films of metallic oxides.

The SNO thin films are fabricated on insulator (001)-oriented $(LaAlO_3)_{0.3}(Sr_2AlTaO_6)_{0.7}$ (LSAT) substrate by a pulsed laser deposition method in vacuo with a back pressure of $10^{-7}$ Torr at 900 °C. A ceramic target composed of $Sr_2Nb_2O_7$ is ablated by a KrF excimer laser (λ = 248 nm) with a repetition rate of 5 Hz and an energy fluence of 0.93-1.14 J/cm$^2$. The crystal structures of SNO (001) thin films are examined by an out-of-plane $2\theta/\omega$ scan and reciprocal space mapping with X-ray diffraction (Rigaku, Smart Lab). The film thickness is determined by a surface profiler (Bruker, Dektak XT) and a deposition rate. The surface uniformity is estimated with an atomic force microscope (AFM). The transport properties are examined by using four-terminal resistance and Hall resistance measurements in a Physical Properties Measurement System (Quantum Design, PPMS) from 2 to 300 K and –2 to 2 T.

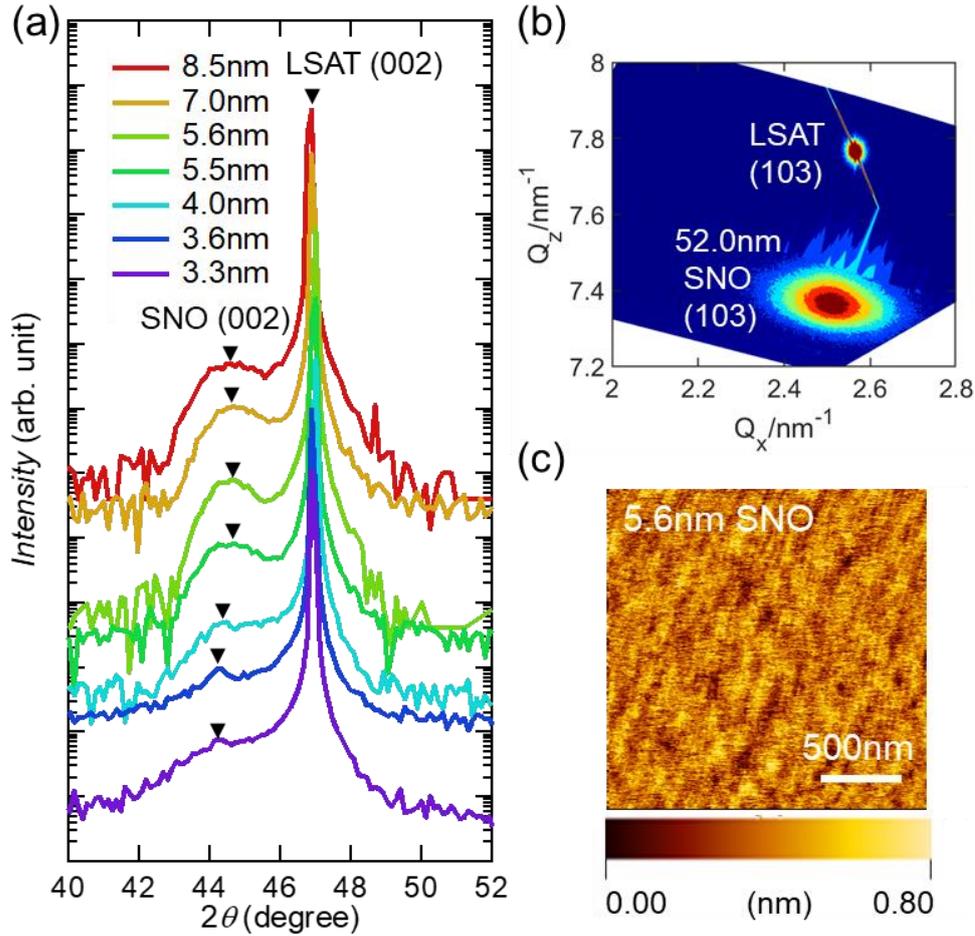

FIG. 1 (a) Out-of-plane 2θ/ω scan of representative SNO/LSAT with various *t* values ranging from 3.3 to 8.5 nm. (b) Reciprocal space mapping on a (103) reflection of SNO/LSAT with *t* = 52.0 nm. (c) AFM surface morphology of SNO/LSAT with *t* = 5.6 nm.

Figure 1(a) shows an out-of-plane 2θ/ω scan of the SNO films on the LSAT substrates (SNO/LSAT) (*a* = 0.3868 nm) with *t* ranging from 3.3 to 8.5 nm. The out-of-plane lattice constant ranged from 4.050 to 4.094 Å, which was always larger than the lattice constant in a pseudocubic approximation (a = 4.023 Å) [27] and is in agreement with a previous report on a SNO film [21]. The crystal structure was directly examined by performing reciprocal space mapping (RSM) on a (103) reflection of SNO/LSAT with *t* = 52.0 nm, as shown in Fig. 1(b). The in-plane and out-of-plane lattice constants were obtained as 3.980 and 4.073 Å, respectively. The cell volume of the SNO film was smaller than that of bulk SNO by approximately 1%, probably indicating small off-stoichiometry variations for the film. The smaller in-plane lattice constant indicates that the film was partly strained by the substrate. Figure 1(c) displays the surface morphology of SNO/LSAT with *t* = 5.6 nm obtained by AFM.

The root mean square (RMS) roughnesses of SNO/LSAT with $t$ ranging from 3.3 to 8.5 nm were always less than 1 nm, indicating that the SNO films had flat surfaces.

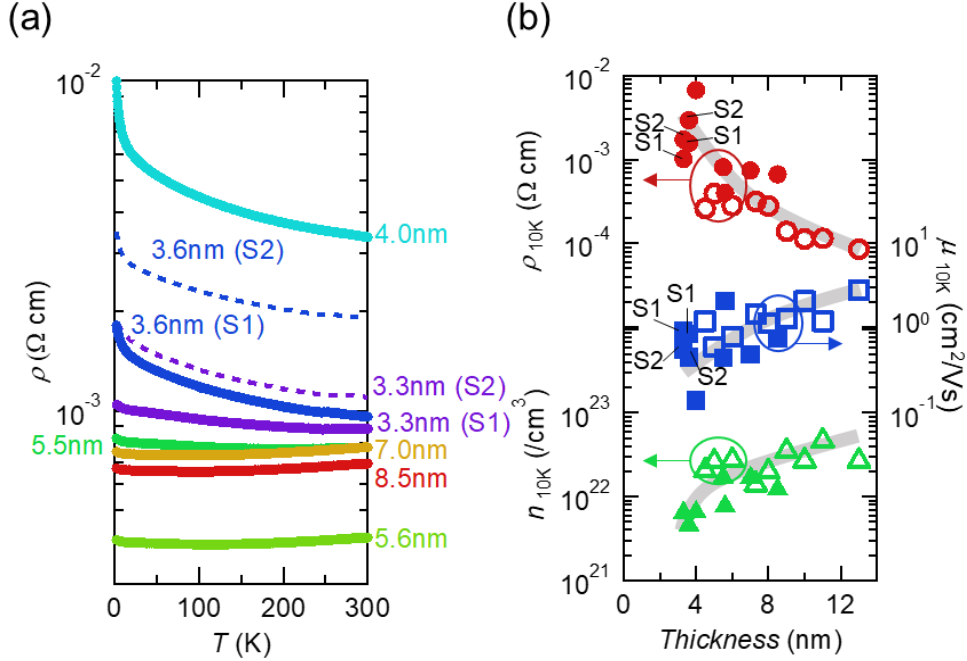

FIG. 2. (a) Temperature dependence of resistivity ($\rho$) for films with various $t$ values ranging from 3.3 to 8.5 nm. (b) $\rho$, carrier density ($n$), and mobility ($\mu$) at 10 K ($\rho_{10K}$, $n_{10K}$, and $\mu_{10K}$) plotted against $t$ for films with $t$ ranging from 3.3 to 13 nm. Solid and open symbols correspond to the data for the films indicated in Fig. 2(a) and others not shown in (a), respectively. The dotted lines are visual references.

We examined the temperature dependence of the transport properties of SNO/LSAT films with various $t$ values. Figure 2(a) shows the resistivity ($\rho$) of the films shown in Fig. 1(a). We observed a metal to insulator (or bad metal) transition (MIT) of the SNO films by reducing $t$. Films thicker than 4.0 nm exhibited metallic behavior with a resistivity minimum at a certain temperature $T_{min}$, which increased with decreasing $t$. In contrast, the films with $t \leq 4.0$ nm exhibited insulating (or bad metal) behavior over the entire temperature range. We obtained the carrier density ($n$) from the Hall resistance measurements and the mobility ($\mu$) from $\rho$ and $n$. In Fig. 2(b), we plotted $\rho$, $n$, and $\mu$ at 10 K (denoted as $\rho_{10K}$, $n_{10K}$, and $\mu_{10K}$, respectively) for the films shown in Fig. 2(a) and others not shown in Fig. 2(a), which are indicated by solid and open symbols, respectively, against $t$ ranging from 3.3 to 13 nm. $\rho_{10K}$ increased by nearly two orders of magnitude with decreasing $t$, while both $n_{10K}$ and $\mu_{10K}$ decreased by nearly one order of magnitude with decreasing $t$. These results indicate that both an increase in disorder and a decrease in carrier density are responsible for the insulating behavior of the thinner films with $t \leq 4.0$ nm. Moreover, we noticed the scattering of the data points of $\rho_{10K}$, $n_{10K}$, and $\mu_{10K}$ in proximity to the individual dotted lines drawn as visual references. This is probably due to the difference in the disorder of the films with a given thickness, such as disorder originating from oxygen vacancies, defects, and surface degradation.

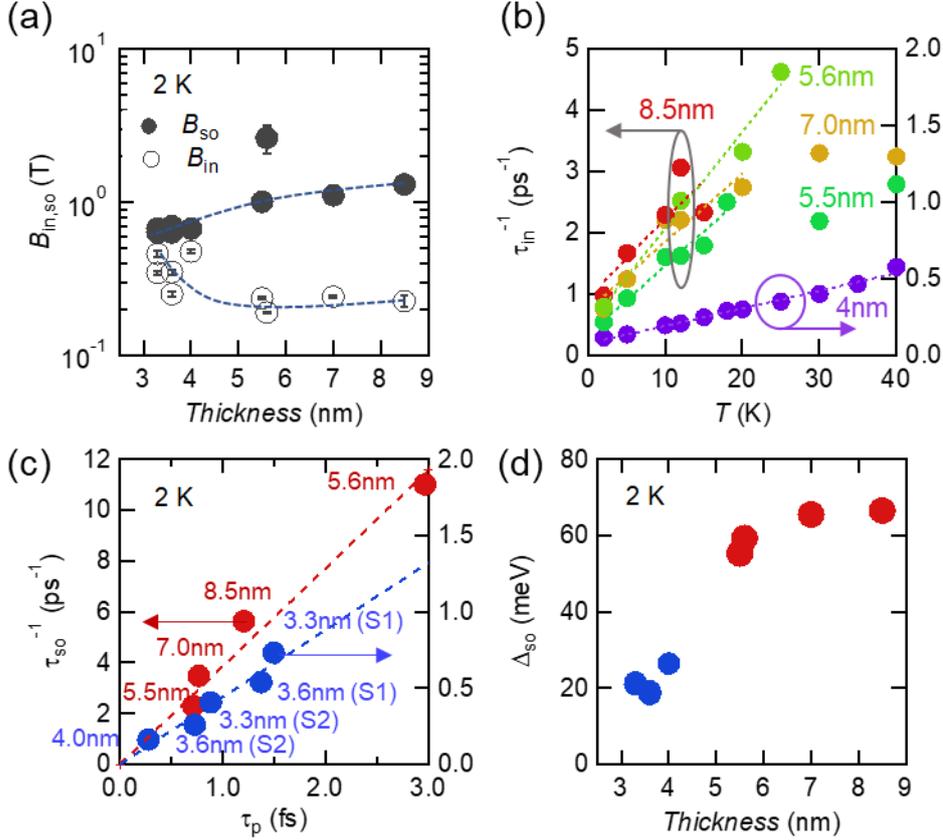

FIG. 3 (a) Effective fields $B_{in}$ (gray solid circles) and $B_{so}$ (gray open circles) extracted from the MR fits at 2 K to Eq. (4) plotted against $t$. The dotted lines are visual references. (b) Inverse of the inelastic scattering time ($\tau_{in}^{-1}$) at 2 K for the films with $t$ ranging from 4 to 8.5 nm as a function of $T$. The dotted straight lines represent the linear fits. (c) Inverse of the spin relaxation time ($\tau_{so}^{-1}$) versus the momentum scattering time ($\tau_p$) at 2 K for the films with $t$ ranging from 5.5 to 8.5 nm (red solid circles) and 3.3 to 4.0 nm (blue solid circles). The dashed straight lines are fits of the data to $\tau_{so}^{-1} \propto \tau_p$. The relation $\tau_{so}^{-1} \propto \tau_p$ clearly indicates that the DP spin relaxation mechanism is dominant in the SNO films rather than the EY spin relaxation mechanism, which follows $\tau_{so} \propto \tau_p$. (d) Energy band splitting due to SOC ($\Delta_{so}$) extracted from the data in (c) using Eq. (9) plotted as a function of $t$.

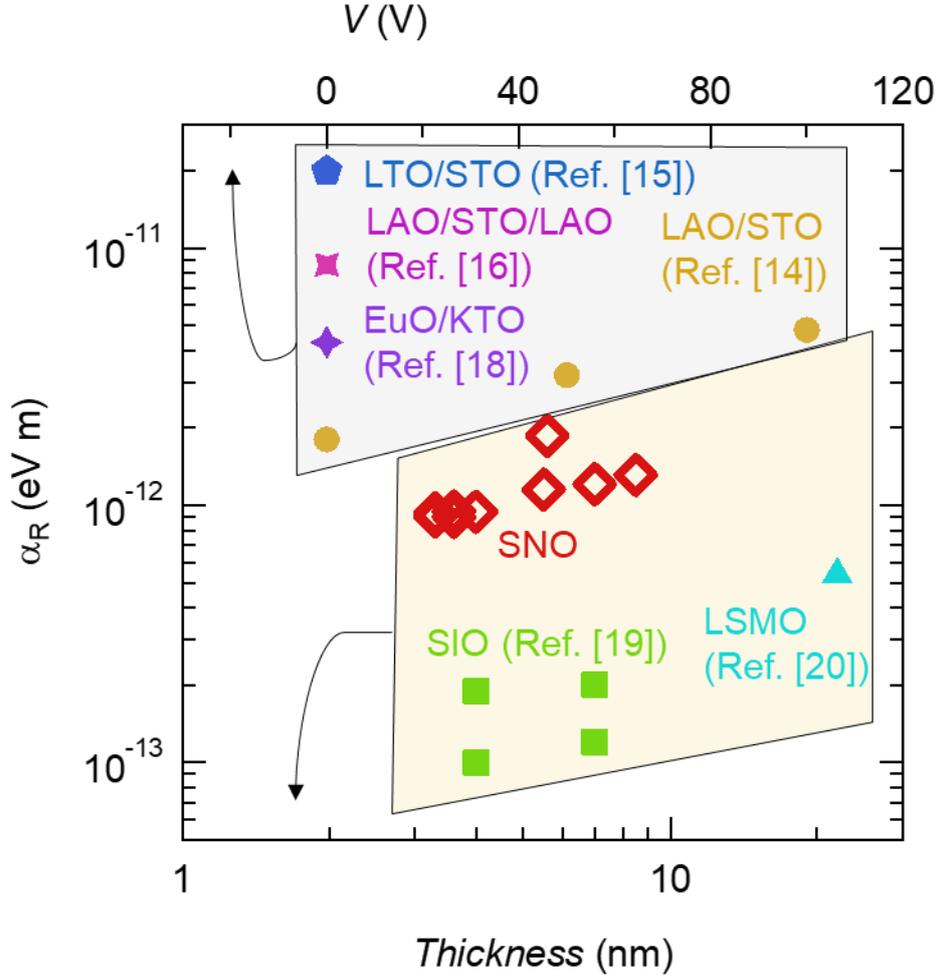

FIG. 4 Rashba parameter ($\alpha_R$) for oxide materials with metallic films (SNO/LSAT, SIO/STO [19], LSMO/STO [20]) and interface heterostructures (LaAlO$_3$(LAO)/STO [14], LaTiO$_3$(LTO)/STO [15], LAO/STO/LAO [16], EuO/KTO [18]).

To estimate the SOC strength, we measured $\rho$ for the films shown in Fig. 1(a) as a function of a perpendicular magnetic field ($B$) at 2 K. These data are shown in the form of a magnetoconductance $\Delta\sigma(B)$ ($\equiv \sigma(B) - \sigma(0)$) in the Supplemental Material, Sec. A and B [28]. All films showed a positive MR, and $\Delta\sigma(B)$ was well reproduced by WAL theory for a weakly disordered film. We fitted $\Delta\sigma(B)$ both with Hikami-Larkin-Nagaoka (HLN) theory, taking into account the EY spin-relaxation mechanism, and Iordanskii-Lyanda-Pikus (ILP) theory, taking into account the DP spin-relaxation mechanism. In HLN theory, MR is described by the following equation [29] [30]:

$$\frac{\Delta\sigma(B)}{\sigma_0} = -\left[\Psi\left(\frac{1}{2} + \frac{(\hbar/4eL_p^2)}{B}\right) - \ln\left(\frac{\hbar/4eL_p^2}{B}\right) - \Psi\left(\frac{1}{2} + \frac{B_{\text{in}} + B_{\text{so}}}{B}\right) + \ln\frac{B_{\text{in}} + B_{\text{so}}}{B}\right], \quad (1)$$

$$L_p = \frac{\hbar k_F \mu}{e}, \quad (2)$$

$$k_F = \sqrt{2 n_s \pi}, \quad (3)$$

where $\sigma_0 = e^2/\pi\hbar$, $e$ is the elementary charge, $\Psi$ is the digamma function, $\hbar$ is the Planck constant, $L_p$ is the momentum scattering length, $B_{in}$ and $B_{so}$ are effective fields related to inelastic and spin-orbit scattering, respectively (for details, see below), $k_F$ is the Fermi wavenumber, and $n_s$ is the sheet carrier density. As shown in Fig. S1 in the Supplemental Material, Sec. A [26], $\Delta\sigma(B)$ at 2 K for films with $t$ = 3.3 and 4.0 nm followed Eq. (1); however, the values of $L_p$ extracted from the fitting were nearly two orders of magnitude larger than those obtained from $k_F$ and $\mu$. Therefore, it is unlikely that the $B$ dependence of the MR is explained by HLN theory taking into account the EY spin-relaxation mechanism.

Next, we focus on ILP theory, which describes the MR with the following equation [31] [32]:

$$\frac{\Delta\sigma(B)}{\sigma_0} = -\left[\frac{1}{2}\Psi\left(\frac{1}{2}+\frac{B_{in}}{B}\right) - \frac{1}{2}\ln\frac{B_{in}}{B} - \Psi\left(\frac{1}{2}+\frac{B_{in}+B_{so}}{B}\right) + \ln\frac{B_{in}+B_{so}}{B} - \frac{1}{2}\Psi\left(\frac{1}{2}+\frac{B_{in}+2B_{so}}{B}\right) + \frac{1}{2}\ln\frac{B_{in}+2B_{so}}{B}\right], \quad (4)$$

As shown in Fig. S2 in the Supplemental Material, Sec. B [26], $\Delta\sigma(B)$ at 2 K followed Eq. (4) for films with $t$ ranging from 3.3 to 8.5 nm. Figure 3(a) shows the $t$ dependence of $B_{in}$ and $B_{so}$ deduced from the fitting by Eq. (4). It was found that $B_{so}$ was always larger than $B_{in}$, indicating that spin-orbit scattering was dominant over inelastic scattering. With a decrease in $t$ to 5.5 nm, $B_{so}$ decreased gradually, while $B_{in}$ remained nearly constant. With a further decrease in $t$, $B_{in}$ increased and was comparable to $B_{so}$ for $t \leq 4.0$ nm. This result implies that inelastic scattering was enhanced as the system entered the insulating (bad metal) regime, and then the contribution of inelastic scattering was comparable to that of spin-orbit scattering.

Then, we examined the three characteristic times (inelastic scattering time ($\tau_{in}$), $\tau_{so}$, and $\tau_p$) for the SNO films. $B_{in}$ and $B_{so}$ are related to an inelastic scattering length (time) $L_{in}$ ($\tau_{in}$) and a spin-orbit scattering length (time) $L_{so}$ ($\tau_{so}$), respectively, with the following equations:

$$B_{in,so} = \frac{\hbar}{4eL^2_{in,so}}, (5)$$

$$L_{in,so} = \sqrt{D\tau_{in,so}}, \quad (6)$$

$$D = \frac{1}{2}v_F^2 \tau_p, \quad (7)$$

$$\tau_p = \frac{\mu m}{e}, \quad (8)$$

where $D$ is the diffusion constant, $v_F$ is the Fermi velocity, and $m$ is the effective mass. We employed $m = 2.76 m_0$ [22] for all films, where $m_0$ is the free electron mass. Figure

3(b) displays the inverse of the inelastic scattering time ($\tau_{in}^{-1}$) as a function of $T$ for various films. The linear relation between $\tau_{in}^{-1}$ and $T$ was observed and was well explained assuming that electron-electron scattering was dominant [33] [34].

Figure 3(c) shows $\tau_{so}^{-1}$ as a function of $\tau_p$ for films $t \geq 5.5$ nm (metallic films, red solid circles) and those with $t \leq 4.0$ nm (insulating films, blue solid circles). For both the metallic and insulating films, $\tau_{so}^{-1}$ was proportional to $\tau_p$, verifying that the main spin relaxation mechanism was DP-type spin relaxation ($\tau_{so} \propto \tau_p^{-1}$) instead of EY-type spin relaxation ($\tau_{so} \propto \tau_p$). Note that the slope of the red dashed line for the metallic films is larger than that of the blue dashed line for the insulating films. The abrupt decrease in $\tau_{so}^{-1}$ near MIT was also reported in ultrathin SIO films with $R_s$ = 3-4 k$\Omega$ below quantum resistance $R_Q$ = 25 k$\Omega$, while $\tau_{so}$ was governed by the EY mechanism of SIO films [35]. For SNO, the $R_s$ of films with $t$ = 3.3-4.0 nm ranged from 3 k$\Omega$ to 25 k$\Omega$, and $L_p$ ranged from 0.12 to 0.75 nm, which are close to the value of the lattice constant. Therefore, one explanation for the sudden reduction of $\tau_{so}^{-1}$ is a change in the band structure at the MIT. According to the DP spin relaxation mechanism, the relation between $\tau_{so}^{-1}$ and $\tau_p$ is expressed as [20] [32] [36] [37] [38]:

$$\tau_{so}^{-1} = \frac{1}{2}(\frac{\Delta_{so}}{\hbar})^2 \tau_p, \quad (9)$$

where $\Delta_{so}$ is the energy band splitting due to SOC and $\alpha_R$ is the Rashba parameter. $\Delta_{so}$ was estimated with Eq. (9) and plotted against $t$ in Fig. 3(d) with red and blue circles, corresponding to films in the metallic and insulating regimes, respectively. Thus, the reduction in $\tau_{so}^{-1}$ is related to the decrease in $\Delta_{so}$ by a factor of approximately 3 at the MIT.

Figure 4 shows $\alpha_R$ at 2 K as a function of $t$. In the metallic regime, $\Delta_{so}$ is related to $k_F$ and $\alpha_R$ with the following expression by assuming a parabolic band structure:

$$\Delta_{so} = 2 k_F \alpha_R, \quad (10).$$

We employed $k_F$ calculated from Eq. (3) using an experimental value of $n_s$, and we obtained $\alpha_R$ from Eq. (10) for films of all thicknesses, although $k_F$ is not a good parameter in the insulator regime. Thus, $\alpha_R$ for films with $t \leq 4.0$ nm is not a very reliable value. The values of $\alpha_R$ for our SNO ultrathin films for all thicknesses were larger than those reported for other ultrathin films of metallic oxides, SrIrO$_3$ (SIO) (green solid squares) and La$_{2/3}$Sr$_{1/3}$MnO$_3$ (LSMO) (light blue solid triangles), and comparable to or smaller than the values reported for oxide heterointerfaces [14] [15] [16] [18], which are on the order of $10^{-12}$ -$10^{-11}$ eVm. It is expected that $\alpha_R$ can be enhanced by reducing the film thickness, while the MIT at the ultrathin film prevents us from seeing this enhancement. Therefore, we expect that, combined with improved fabrication and a sophisticated carrier doping technique, SNO metallic ultrathin films with a few monolayers will be a promising material system for practical applications in Rashba-based spintronics devices.

In summary, to reveal the spin relaxation mechanism of SNO ultrathin films, the transport properties of a series of films with different *t* were measured on both sides of the MIT. The MR at 2 K for the films with various *t* and disorder was analyzed using WAL theory, and it was found that $\tau_{so} \propto \tau_p^{-1}$. This result verified the DP-type spin relaxation mechanism, indicating the dominant Rashba effect. The values of $\alpha_R$ were on the order of $1\times10^{-12}$ eVm, which were largest in the values reported for other ultrathin films of metallic oxides.

This work was supported in part by JSPS KAKENHI (Grant Nos. 21H01038 and 19H02798) and CREST-JST (Grant No. JPMJCR15Q2).

**Supplementary Material**

**Large Rashba parameter for 4*d* strongly correlated perovskite oxide SrNbO₃ ultrathin films**

Hikaru Okuma*, Yumiko Katayama, and Kazunori Ueno
Department of Basic Science, University of Tokyo,
Meguro, Tokyo 153-8902, Japan
*hikaruokuma613@g.ecc.u-tokyo.ac.jp

A. Fitting to HLN theory taking into account the EY spin-relaxation mechanism

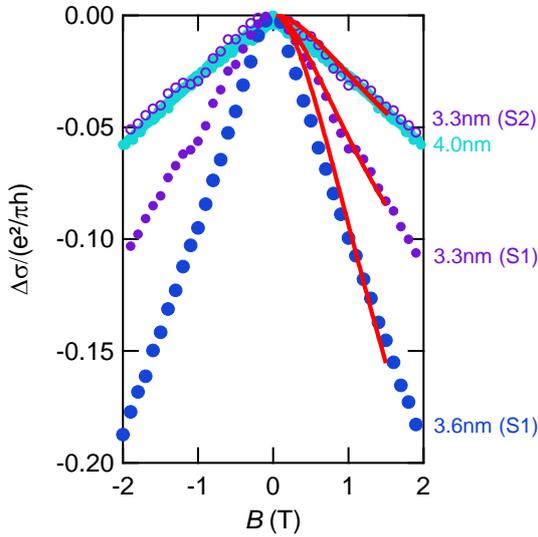

FIG. S1. Δσ=σ(B)-σ(0) in units of $e^2/\pi\hbar$ measured in the perpendicular magnetic field $B$ for films with $t$ ranging from 3.3 nm to 4.0 nm at 2 K. The red solid lines are fits to the data based on HLN theory [Eq. (1)].

Figure S1 shows Δσ(B) at 2 K and the fitting of the data to HLN theory [Eq. (1)] taking into account the EY spin-relaxation mechanism. Δσ(B) for films with $t$ = 3.3 (S1), 3.3 (S2), and 4.0 nm followed Eq. (1). However, Δσ(B) for film with $t$ = 3.6 nm (S1) deviated from the fit over a low $B$ range, and for films with $t$ = 5.5-8.5 nm and 3.6 nm (S2), Δσ(B) did not follow Eq. (1) (not shown here). Thus, the values of $L_p$ deduced from the fitting of the data for the films with $t$ = 3.3 (S1), 3.3 (S2), 3.6 (S1), and 4.0 nm were 20.4, 17.9, 19.4, and 17.5 nm, and those obtained from $k_F$ and $\mu$ were 0.75, 0.44, 0.60, and 0.15 nm, respectively. Notably, the values of $L_p$ deduced from the fitting were nearly two orders of magnitude larger than those obtained from $k_F$ and $\mu$. Therefore, it is unlikely that the $B$ dependence of the MR is explained by HLN theory taking into account the EY spin-relaxation mechanism.

B. Fitting to ILP theory taking into account the DP spin-relaxation mechanism

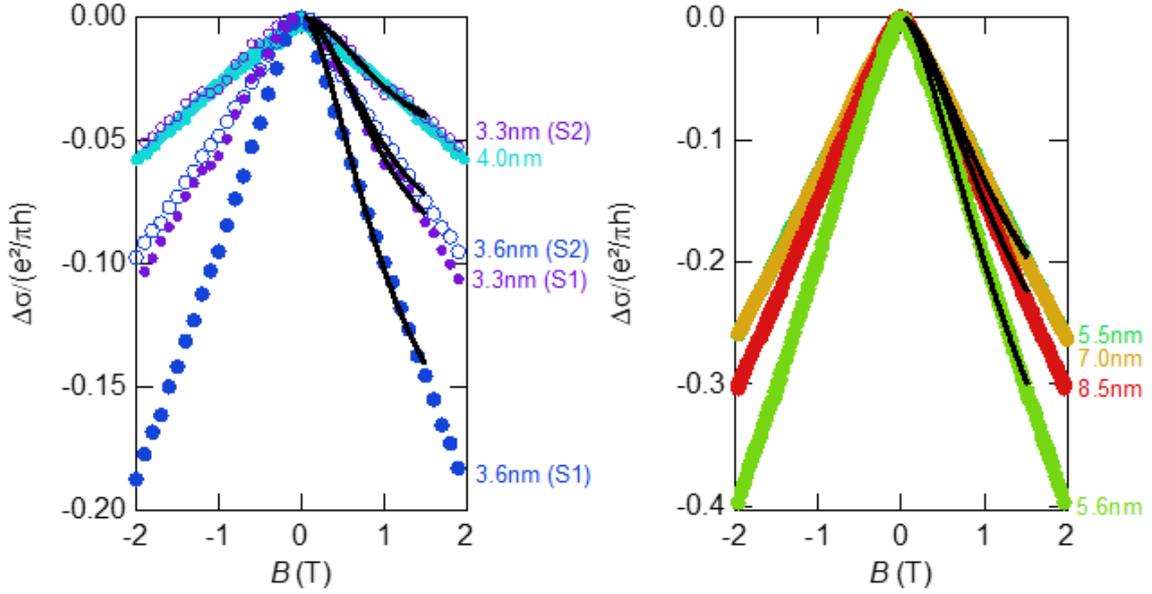

FIG. S2. $\Delta\sigma=\sigma(B)-\sigma(0)$ in units of $e^2/\pi\hbar$ measured in the perpendicular magnetic field $B$ for films with $t$ ranging from 3.3 to 4.0 (left) and 5.5 nm to 8.5 nm (right) at 2 K. The black solid lines are fits to the data based on ILP theory [Eq. (4)].

Figure S2 shows $\Delta\sigma(B)$ at 2 K and the fitting of the data to Iordanskii-Lyanda-Pikus (ILP) theory [Eq. (4)] taking into account the DP spin-relaxation mechanism. $\Delta\sigma(B)$ follows Eq. (4) for films with $t$ ranging from 3.3 to 8.5 nm.

## C. Film thickness dependence of the momentum scattering length

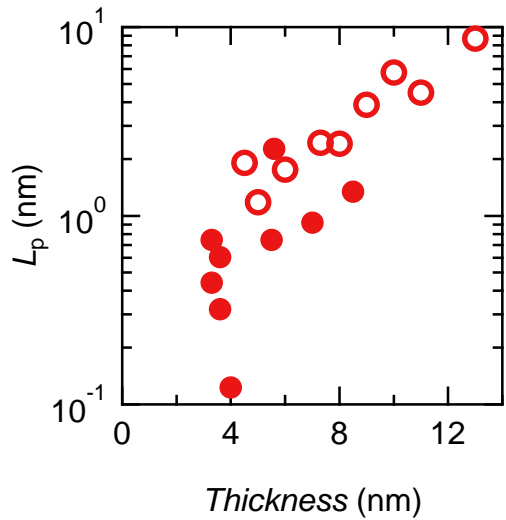

FIG. S3. $L_p$ plotted against $t$ for films with $t$ ranging from 3.3 to 13 nm. Solid and open symbols correspond to the data for the films indicated in Fig. 2(a) and others not shown in (a), respectively.

Figure S3 shows the $L_p$ for the films shown in Fig. 2 (a) and others not shown in Fig. 2(a), which are indicated by solid and open symbols, respectively, against $t$ ranging from 3.3 to 13 nm. $L_p$ decreased by nearly two orders of magnitude with decreasing $t$. Films thicker than 4 nm exhibited metallic behavior, and $L_p$ ranged from 0.75 to 8.69 nm. In contrast, for the film with $t \leq 4$ nm, $L_p$ ranged from 0.12 to 0.75 nm, which is close to the value of the lattice constant. Therefore, films with $t \leq 4$ nm exhibited insulating or poor metallic properties, which is consistent with the $\rho$-$T$ curves shown in Fig. 2(a).